\documentclass[preprint,12pt]{elsarticle}

\usepackage{color,longtable}

\definecolor{mygreen}{rgb}{0,1,0}
\definecolor{mygreen}{rgb}{0,.75,0}

\definecolor{mycyan}{cmyk}{1,0,0,0}
\definecolor{mycyan}{cmyk}{.8,.15,0,0}
\definecolor{mycyan}{cmyk}{.8,.55,0,0}

\definecolor{mymagenta}{cmyk}{0,1,0,0}
\definecolor{mymagenta}{cmyk}{.15,1,0,0}

\usepackage{amssymb}

\biboptions{compress}

\newcounter{bla}

\journal{Computer Physics Communications}

\newcommand{\HAWK}{{\sc HAWK}}

\def\reffi#1{\mbox{Fig.~\ref{#1}}}

\def\refse#1{\mbox{Section~\ref{#1}}}

\def\citere#1{\mbox{Ref.~\cite{#1}}}
\def\citeres#1{\mbox{Refs.~\cite{#1}}}

\def\mathswitch#1{\relax\ifmmode#1\else$#1$\fi}
\def\mathswitchr#1{\relax\ifmmode{\mathrm{#1}}\else$\mathrm{#1}$\fi}
\def\mathswitchit#1{\relax\ifmmode{#1}\else$#1$\fi}

\newcommand{\PV}{\mathswitch V}
\newcommand{\PW}{\mathswitchr W}
\newcommand{\PZ}{\mathswitchr Z}

\newcommand{\Pg}{\mathswitchr g}
\newcommand{\PH}{\mathswitchr H}

\newcommand{\Pne}{\mathswitch \nu_{\mathrm{e}}}
\newcommand{\Pane}{\mathswitch \bar\nu_{\mathrm{e}}}

\newcommand{\Pl}{\mathswitch l}

\newcommand{\Pnl}{\mathswitch {\nu_\Pl}}
\newcommand{\Pnlbar}{\mathswitch {\overline{\nu}_\Pl}}

\newcommand{\Pb}{\mathswitchr b}

\newcommand{\Pt}{\mathswitchr t}
\newcommand{\Pp}{\mathswitchr p}
\newcommand{\Ppbar}{\overline{\Pp}}

\newcommand{\Pmum}{\mathswitchr {\mu^-}}
\newcommand{\Pmup}{\mathswitchr {\mu^+}}

\newcommand{\MW}{\mathswitch {M_\PW}}

\newcommand{\MZ}{\mathswitch {M_\PZ}}
\newcommand{\MH}{\mathswitch {M_\PH}}

\newcommand{\Mt}{\mathswitch {m_\Pt}}

\newcommand{\sw}{\mathswitch {s_{\PW}}}
\newcommand{\cw}{\mathswitch {c_{\PW}}}

\newcommand{\GeV}{\unskip\,\mathrm{GeV}}

\newcommand{\kt}{k_{\mathrm{T}}}

\newenvironment{cpcdescription}
   {\begin{description}%
   \setlength{\itemsep}{2ex}}%
   {\end{description}}

\newlength{\colonewidth}
\newlength{\parwidth}

\newcommand{\cpcitemtable}[2]
{\settowidth{\colonewidth}{#1}\setlength{\parwidth}{\textwidth}%
\addtolength{\parwidth}{-\leftmargin}%
\addtolength{\parwidth}{-\colonewidth}\addtolength{\parwidth}{-2em}%
\nobreak
\begin{flushleft}%
\vspace{-1.5ex}
\begin{tabular}[l]{@{}p{\colonewidth}@{ }c@{ }p{\parwidth}}%
#2 
\end{tabular}%
\vspace{-1.ex}
\end{flushleft}%
\ignorespaces}

\newcommand{\cpctable}[2]
{\settowidth{\colonewidth}{#1}\setlength{\parwidth}{\textwidth}%
\addtolength{\parwidth}{-\leftmargin}%
\addtolength{\parwidth}{-\colonewidth}\addtolength{\parwidth}{-2em}%
\nobreak
\begin{longtable}[l]{@{\qquad}p{\colonewidth}@{ }c@{ }p{\parwidth}}%
#2 
\end{longtable}%
\noindent\ignorespaces}

\newcommand{\pt}{p_\mathrm{T}}

\def\draftdate{\relax}
\def\mda{\relax}
\def\mua{\relax}
\def\mla{\relax}
\def\draft{
\def\thtystars{******************************}
\def\sixtystars{\thtystars\thtystars}
\typeout{}
\typeout{\sixtystars**}
\typeout{* Draft mode!
         For final version remove \protect\draft\space in source file *}
\typeout{\sixtystars**}
\typeout{}
\def\draftdate{\today}
\def\mua{\marginpar[\boldmath\hfil$\uparrow$]%
                   {\boldmath$\uparrow$\hfil}%
                    \typeout{marginpar: $\uparrow$}\ignorespaces}
\def\mda{\marginpar[\boldmath\hfil$\downarrow$]%
                   {\boldmath$\downarrow$\hfil}%
                    \typeout{marginpar: $\downarrow$}\ignorespaces}
\def\mla{\marginpar[\boldmath\hfil$\rightarrow$]%
                   {\boldmath$\leftarrow $\hfil}%
                    \typeout{marginpar: $\leftrightarrow$}\ignorespaces}
\def\Mua{\marginpar[\boldmath\hfil$\Uparrow$]%
                   {\boldmath$\Uparrow$\hfil}%
                    \typeout{marginpar: $\Uparrow$}\ignorespaces}
\def\Mda{\marginpar[\boldmath\hfil$\Downarrow$]%
                   {\boldmath$\Downarrow$\hfil}%
                    \typeout{marginpar: $\Downarrow$}\ignorespaces}
\def\Mla{\marginpar[\boldmath\hfil$\Rightarrow$]%
                   {\boldmath$\Leftarrow $\hfil}%
                    \typeout{marginpar: $\Leftrightarrow$}\ignorespaces}
\overfullrule 5pt
\oddsidemargin -15mm
\marginparwidth 29mm
}

\begin{document}

\begin{frontmatter}



\title{\HAWK\ 2.0: A Monte Carlo program for Higgs production in 
vector-boson fusion and Higgs strahlung at hadron colliders}


\author[a]{Ansgar Denner}
\author[b]{Stefan Dittmaier}
\author[c]{Stefan Kallweit}
\author[d]{Alexander M\"uck\corref{author}}

\cortext[author] 
{Corresponding author.\\\textit{E-mail address:} mueck@physik.rwth-aachen.de}
\address[a]{Universit\"at W\"urzburg, Institut f\"ur Theoretische Physik und Astrophysik, \\
D-97074 W\"urzburg, Germany}
\address[b]{Albert-Ludwigs-Universit\"at Freiburg, Physikalisches Institut, \\
D-79104 Freiburg, Germany}
\address[c]{Johannes Gutenberg-Universit\"at Mainz, 
Institut f\"ur Physik,\\ PRISMA Cluster of Excellence,\\
D-55099 Mainz, Germany}
\address[d]{RWTH Aachen University, Institut f\"ur Theoretische Teilchenphysik und Kosmologie, \\
D-52056 Aachen, Germany}

\begin{abstract}
The Monte Carlo integrator \HAWK\ provides precision predictions for Higgs production
at hadron colliders in vector-boson fusion and Higgs strahlung, i.e.\ in production processes
where the \underline{H}iggs boson is \underline{A}ttached to \underline{W}ea\underline{K} bosons. 
The fully differential predictions include
the full QCD and electroweak next-to-leading-order corrections. Results are computed as
integrated cross sections and as binned distributions for important hadron-collider 
observables. 
\end{abstract}


\end{frontmatter}



\noindent
{\bf PROGRAM SUMMARY}\\

\begin{small}
\noindent
{\em Manuscript Title:}                                       
\HAWK\ 2.0: A Monte Carlo program for Higgs production in 
vector-boson fusion and Higgs strahlung at hadron colliders   \\[1ex]
{\em Authors:}
Ansgar Denner, 
Stefan Dittmaier,
Stefan Kallweit,
Alexander M\"uck\\[1ex]
{\em Program Title:} 
\HAWK, version 2.0                                          \\[1ex]
{\em Journal Reference:}                                      \\[1ex]
{\em Catalogue identifier:}                                   \\[1ex]
{\em Licensing provisions:}
none                                   \\[1ex]
{\em Programming language:}
Fortran 77, Fortran 90                                   \\[1ex]
{\em Computer:}
Any computer with a Fortran 90 compiler                          \\[1ex]
{\em Operating system:}
Linux, Mac OS                                       \\[1ex]
{\em RAM:} less than 1 GB                                              \\[1ex]
{\em Classification:}
4.4 Feynman Diagrams,
11.1 General, High Energy Physics and Computing,
11.2 Phase Space and Event Simulation.                 \\[1ex]
{\em External routines/libraries:} 
LHAPDF\\
(https://www.hepforge.org/downloads/lhapdf).      \\[1ex]
{\em Nature of problem:}
Precision calculation of cross sections and differential distributions for Higgs-boson
production in vector-boson fusion and Higgs strahlung at the LHC as described 
in~\citeres{Ciccolini:2007jr2,Ciccolini:2007ec2,Denner:2011id2}.\\[1ex]
{\em Solution method:}
Multi-channel Monte Carlo integration of perturbative matrix elements including higher-order
QCD and electroweak corrections which are based on a Feynman-diagrammatic calculation.
   \\[1ex]
{\em Restrictions:}
For vector-boson fusion, only the Higgs-boson decay into a pair of
massless singlets is supported. For Higgs strahlung, decay products
are not supported, while the Higgs boson can be off shell.\\[1ex]
{\em Running time:}
Meaningful results can be obtained within a few hours on a single core. The statistical 
uncertainty of the results typically improves with the square root of the run time.

%
\end{small}%


\section{Introduction}

After the discovery of the Higgs boson~\cite{Chatrchyan:2012ufa,Aad:2012tfa}, 
precision predictions for its production
channels are of paramount importance. The
measurement of the Higgs properties is and will continue to be one of the most
important tasks in the physics programme at CERN's Large Hadron 
Collider (LHC)~\cite{Chatrchyan:2013mxa,Chatrchyan:2012jja,Aad:2013wqa,Aad:2013xqa}. Searching
for possible deviations from its properties as predicted by the Standard Model (SM) of particle
physics requires precise predictions for the underlying 
measurements~\cite{Dittmaier:2011ti,Dittmaier:2012vm,Heinemeyer:2013tqa}. While the
total Higgs production rate is dominated by the gluon-fusion process, Higgs
production in vector-boson fusion (VBFH) and in association with a weak boson, also
known as Higgs strahlung, play an important role in unveiling the nature of the Higgs boson. In
this work, we present the Monte Carlo program \HAWK~\cite{hawk} which provides precision
predictions for these production channels, which are based 
on~\citeres{Ciccolini:2007jr,Ciccolini:2007ec,Denner:2011id}.
In particular, the electroweak (EW) corrections contained in the cross-section
predictions for VBFH and VH compiled by the LHC Higgs Cross Section Working Group 
(LHCHXSWG)~\cite{Chatrchyan:2013mxa,Chatrchyan:2012jja,Aad:2013wqa,Aad:2013xqa}
were obtained with the \HAWK\ program.

\begin{figure}
\begin{center}
\includegraphics[width=10cm]{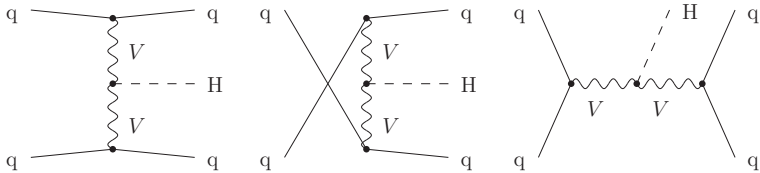}
\end{center}
\caption{\label{fi:diagrams} Leading-order diagrams for VBFH production. The diagrams
correspond to the $t$-channel (left),
$u$-channel (middle), and $s$-channel (right).}
\end{figure}

Higgs production in vector-boson fusion is experimentally accessed by
measuring the Higgs decay products in association with two forward
tagging jets having a large rapidity separation. Applying the
corresponding cuts, as suggested in
\citeres{Barger:1994zq,Rainwater:1997dg,Rainwater:1998kj,Rainwater:1999sd,DelDuca:2006hk},
the cross section is dominated by tree-level diagrams where two quark
lines exchange a weak vector boson in the $t$~channel and the
Higgs boson is emitted by the weak boson. The corresponding
diagrams are shown in \reffi{fi:diagrams}. At the same order in the
weak coupling constant there are also $s$-channel diagrams. Their
contribution to the cross section is suppressed if vector-boson fusion
cuts are applied.
In fact, the $s$-channel contribution corresponds to what is
usually called the Higgs-strahlung process where the weak boson decays hadronically
($\Pp\Pp\to \PH\PV\to \PH+2\,\mathrm{jets}$). \HAWK\ includes the full set of
diagrams ($s$, $t$, and $u$ channels), 
all interferences, and the corresponding next-to-leading 
order~(NLO) QCD and EW corrections~\cite{Ciccolini:2007jr,Ciccolini:2007ec}. 
Hence, the
distinction between vector-boson fusion and Higgs strahlung with a hadronically
decaying weak boson can be done in a completely physical way employing the
corresponding experimental cuts. 

The Higgs-strahlung process $\Pp\Pp\to \PH\PV$ is experimentally
promising when the weak boson $\PV$ decays leptonically. The leptons provide
trigger objects and allow for background suppression. The processes 
$\Pp\Pp\to \PH\PZ\to \PH \Pl^+\Pl^-$, $\Pp\Pp\to \PH\PZ\to \PH \Pnlbar\Pnl$,
$\Pp\Pp\to \PH\PW^+\to \PH \Pl^+\Pnl$, and 
$\Pp\Pp\to \PH\PW^-\to \PH \Pnlbar\Pl^-$ are also calculated by HAWK
including the complete NLO QCD and EW corrections in the SM~\cite{Denner:2011id}.

The theory predictions for the considered processes
with respect to higher orders in QCD corrections are quite advanced.
The total cross section for vector-boson fusion 
is known to NNLO QCD accuracy in the so-called DIS 
approximation~\cite{Bolzoni:2010xr,Bolzoni:2011cu}. 
Differential vector-boson fusion at NLO QCD 
accuracy~\cite{Spira:1997dg,Han:1992hr,Figy:2003nv,Baglio:2014uba}
was matched to parton showers~\cite{Heinemeyer:2013tqa,Nason:2009ai,Alwall:2014hca}. 
The NLO QCD and EW corrections are also available in the alternative Monte Carlo program 
VBFNLO~\cite{Baglio:2014uba,Figy:2010ct}, however, restricted to the
$t$-channel approximation which is sufficient only if vector-boson fusion~(VBF) cuts are applied.

The Higgs-strahlung process was calculated in NNLO 
QCD~\cite{Brein:2003wg,Hamberg:1990np,Brein:2011vx,Brein:2012ne}, and the dominant parts
are now also available in a fully differential 
form~\cite{Ferrera:2011bk,Ferrera:2014lca}. The NLO QCD 
prediction~\cite{Han:1991ia} is known for a long time, and is 
available in public programs~\cite{vv2h,mcfm}. It was also matched
to parton showers~\cite{Hamilton:2009za} and merged {with} matched NLO calculations 
for higher jet multiplicities~\cite{Luisoni:2013cuh}. The EW corrections to inclusive 
cross sections for Higgs strahlung were presented in~\citere{Ciccolini:2003jy}.
The loop-induced process $\Pg\Pg\to \PZ\PH$ and the corresponding NLO QCD
corrections were calculated in the large-$\Mt$ approximation in \citere{Altenkamp:2012sx}.

\HAWK\ provides the state-of-the-art tool to predict fully differential EW
corrections~\cite{Ciccolini:2007jr,Ciccolini:2007ec,Denner:2011id} which can be combined
via differential reweighting with the most advanced QCD predictions. The usual
hadron-collider cuts are available and can be imposed via a simple input
structure. The standard distributions are also available and provided via text
files which are easy to process. Additional user-defined cuts and distributions
are easy to implement, and support from the authors is provided on request.

Within \HAWK, the complex-mass scheme~\cite{Denner:1999gp,Denner:2005fg} 
is employed to calculate the EW
corrections to the Higgs-strahlung processes with potentially resonant weak 
bosons. \HAWK\ implements matrix elements which are calculated in the 
traditional Feynman-diagrammatic approach~\cite{Kublbeck:1990xc,Hahn:2000kx,Hahn:1998yk,Dittmaier:1998nn} 
employing advanced techniques for tensor reduction to standard scalar integrals 
with complex masses~\cite{Passarino:1979jh,Denner:2002ii,Denner:2005nn,Denner:2010tr}.
It employs Monte Carlo techniques 
for phase-space integration which are detailed in~\citeres{Dittmaier:2002ap,Denner:2002cg}.
The combination of real and virtual pieces is based on dipole 
subtraction~\cite{Catani:1996vz,Dittmaier:2008md,Dittmaier:1999mb}.

\HAWK\ can be employed for proton--proton colliders like the LHC, but also 
proton--anti-proton colliders. 
As part of the full NLO EW corrections, \HAWK\ provides predictions
for partonic channels with incoming photons. Within 
\HAWK, the decay of the Higgs boson into a pair of massless gauge singlets
is supported for the VBFH channel. The Higgs boson can be produced on-shell,
or different options for an off-shell Higgs propagator can be used.
Effects of anomalous HZZ/HWW couplings can be included in the predictions.
For VBFH, we also include leading heavy-Higgs-boson effects at 
two-loop order which, however, are negligible for the observed Higgs boson
at 125 GeV. 

This work is structured as follows.
In \refse{se:details}, we {highlight some details of 
\HAWK\ in connection with the underlying physics}. In \refse{se:usage}, 
all the necessary information for the user is given
how to run and use the \HAWK\ program. 
Along with the practical information on all the user-defined
input, we give some details on the program itself. \refse{se:output} introduces 
the way \HAWK\ provides the computed results. In \refse{se:conclusion}, we 
make some concluding remarks.

\section{{EW corrections and off-shell effects in \HAWK}}
\label{se:details}

Specific strengths of the predictions provided by the \HAWK\ program 
concern the careful treatment of EW corrections and off-shell effects{, as 
discussed in the following sections.}

\subsection{$\alpha_{G_\mathrm{F}}$ scheme}
 {The $\alpha_{G_\mathrm{F}}$ scheme is used as input-parameter scheme.}
 The electromagnetic coupling constant is derived from the Fermi constant {according to
 \[
 \alpha_{G_\mathrm{F}} = \frac{\sqrt{2}G_\mathrm{F}\MW^2}{\pi}\left(1-\frac{\MW^2}{\MZ^2}\right) \, .
 \]
This} procedure takes into account some higher-order effects already at
tree level.

\subsection{Complex-mass scheme and input masses}
Gauge-boson resonances are treated using the complex-mass
scheme~\cite{Denner:2005fg}, i.e.\ all {corresponding} decay and
off-shell effects are supported in NLO accuracy in the full phase
space. The program reads the on-shell masses and widths and translates
them internally to the pole masses and widths
{(see~\refse{se:Input_parameters})}.  The latter are then used in
propagators, the complex weak mixing angle, and other derived
couplings.

\subsection{{Final-state photons}}
Both in the VBFH process and in $\PV\PH$ production, the potential appearance of 
an additional photon (as part of the real photonic EW corrections) requires
particular attention in the definition of infrared(IR)-safe observables
{(see also~\refse{se:Recombination})}.

In VBFH production, where two hard jets and a photon appear in the final state of the
real NLO EW correction, IR-safety is achieved in \HAWK\ upon treating the photon as
any other parton in the jet algorithm when a photon--jet pair becomes unresolved. 
In case the photon and a jet are merged, the resulting object is treated as a jet.

In $\PV\PH$ production, where the decay of $\PV$ produces at least one charged lepton,
IR issues occur when the photon becomes collinear to a charged lepton.
If the lepton is a muon, \HAWK\ admits a perfect isolation of the photon
from the muon, leading to final-state-radiation corrections that are enhanced
by a logarithm $\alpha\ln(m_\mu/Q)$ of the muon mass $m_\mu$ over some large scale $Q$.
Technically, the lepton--photon isolation is performed according 
to~\citere{Dittmaier:2008md}.
If the lepton is an electron or positron, collinear photons and electrons/positrons
appear as one electromagnetic shower in the detector, so that photon and lepton are
recombined to a single quasi-particle when they become sufficiently collinear.
After this photon recombination no logarithmically enhanced final-state-radiation effects
remain.

\subsection{Combination of EW corrections from \HAWK\ with external 
QCD-based predictions}
\label{se:combination}
Apart from producing predictions with full NLO QCD+EW accuracy, \HAWK\ can be
used to calculate differential EW correction factors for any one- or more-dimensional
distribution {$\mathrm{d}\sigma/\mathrm{d} \mathcal{O}$, which can be used as differential EW reweighting factors to improve
other QCD-based predictions for VBFH or $\PV\PH$ production that go beyond
NLO accuracy. We suggest to use
\[
  \frac{\mathrm{d}\sigma^{\mathrm{best}}_{\mathrm{QCD}\times\mathrm{EW}}}{\mathrm{d} \mathcal{O}} 
  = \left[1+\delta_{\mathrm{EW}}(\mathcal{O})\right]
    \frac{\mathrm{d}\sigma^{\mathrm{best}}_{\mathrm{QCD}}}{\mathrm{d}\mathcal{O}} +
    \frac{\mathrm{d}\sigma_{\gamma}}{\mathrm{d}\mathcal{O}}\,, 
\]
i.e.\ to assume approximate factorization for QCD and EW corrections. In the above formula,
$\mathrm{d}\sigma^{\mathrm{best}}_{\mathrm{QCD}}$ is the best 
available QCD prediction to be obtained, for example, by one of the QCD calculations
mentioned in the introduction. \HAWK\ provides the differential EW corrections
$\delta_{\mathrm{EW}}(\mathcal{O})$  and the differential cross section $\mathrm{d}\sigma_{\gamma}$
due to photons in the initial state (see \refse{se:Differential}). 
In particular, the total cross sections reported in
\citeres{Dittmaier:2011ti,Dittmaier:2012vm,Heinemeyer:2013tqa} by the LHCHXSWG
were obtained in this way. 
Note that EW corrections provided by \HAWK\ are almost independent of the PDF 
set which is used in the calculation.
Note also that \HAWK\ provides all results in form of binned distributions.
Unweighted events are not available.
}

\subsection{Higgs-boson decays}

The current version of \HAWK\ optionally supports only the Higgs-boson decay
channel into a pair of massless singlets without any correction to this decay. 
The corresponding branching ratio has to be provided as external input
for the specific Higgs-boson decay (see \refse{se:off-shell}). 
Note that the decay into two photons is well approximated 
in \HAWK\ by the singlet decay. Since there are no corrections due to
real radiation to this decay at NLO, 
the (purely virtual) EW corrections can be included by the appropriate $\PH\to\gamma\gamma$ branching ratio.

\section{The usage of \HAWK}
\label{se:usage}

\subsection{\HAWK\ installation}

\HAWK\ is a stand-alone Fortran 77/90 code~\cite{hawk}. No additional libraries are
needed to run \HAWK, apart from the LHAPDF library~\cite{LHAPDF} to employ
up-to-date PDF sets, as discussed below. 
It has been tested under various operating systems, 
including different Linux distributions and MAC OS without problems. It should 
also be compilable with any standard Fortran compilers and has been successfully
tested using GNU Fortran (GCC) 4.1.2, 4.4.7, 4.6.3, 4.7.2, pgf95 10.1-0,
Intel Fortran Compiler 11.1, 13.1, and Openmpi 1.6.4. 

For installation,
{\tt gunzip} and {\tt untar HAWK-2.0.tar.gz} which will unpack into the directory  
{\tt ./HAWK-2.0.} The source files are contained in the {\tt HAWK-2.0/src} directory.
The directory {\tt HAWK-2.0/sampleruns} contains input files and the corresponding 
results of sample runs.
The directories {\tt HAWK-2.0/bin}, {\tt HAWK-2.0/obj}, and {\tt HAWK-2.0/mod}
are for the executable, object files, and module files, respectively. To 
select the compiler or compiler options, the {\tt Makefile} can be edited. To
compile \HAWK, simply go to the {\tt HAWK-2.0} directory and 
issue {\tt make} in the command line to generate the executable {\tt hawk-2.0}. 

For usage of up-to-date parton distributions functions (PDFs), an interface to
the LHAPDF library~\cite{LHAPDF} is provided. 
The LHAPDF library is not part of the
\HAWK\ distribution. In order to use LHAPDF (which is the default option), 
the preprocessor flag {\tt includeLHAPDF} must
be set in the {\tt Makefile}, and the corresponding library must be
provided. This can be done by setting the variable {\tt LHAPDF\_CONFIG} to the
executable {\tt lhapdf-config} of the LHAPDF version that should be used.
Alternatively, the path to the LHAPDF library and, if needed, include
files can be set manually via {\tt LHAPDF} and {\tt LHAPDFFFLAGS}, respectively.
For usage without LHAPDF, only the code and tables for two PDF sets 
(MRST2004QED and CTEQ6) are included in the distribution.

\subsection{\HAWK\ execution}

To run \HAWK, an input file from the standard input is needed
(see, however, the remarks for parallel execution using mpi in \refse{se:mpi}).
The program can be executed using:\\[1ex]  
\mbox{} \hspace{1cm}
{\tt ./hawk-2.0 < inputfile}\\[1ex]
Output will be written to standard output. If an output file is specified in the 
input file, it will be used to store the central results.

All input is delivered via the inputfile. This has to be specified
via\linebreak standard input, otherwise HAWK does not start.
Its general format\linebreak can be seen from the default inputfile {\tt input\_default}
in the directory\linebreak {\tt HAWK-2.0/sampleruns} and the input files of the subdirectories.
While the sample input files specify all relevant parameters, it is
sufficient to specify those values that differ from the default.
As a general remark, do not forget the ``d0" after ``double precision" 
quantities.

When using LHAPDF, the relevant files for the PDFs should be
in the directory {\tt PDFsets} of the LHAPDF installation (see \refse{se:PDF}). 
When using the included PDF sets, the corresponding table files
({\tt qed6-10gridp.dat} for MRST2004QED or {\tt cteq6m.tbl}, {\tt cteq6l.tbl},
{\tt cteq6l1.tbl} for CTEQ6) must be available in the working directory where
\HAWK\ is executed.

\subsection{\HAWK\ input}
\label{se:input}

The detailed description of \HAWK\ in the following sections follows the structure of the 
default input file {\tt input\_default} which is provided with the \HAWK\ distribution. All supported
options for a given user input are detailed, and the underlying physics is 
briefly explained.

\subsubsection{Global parameters}

\begin{cpcdescription}
\item[{\bf {\tt outputfile:}}] a character string that specifies the name of the output file.
\cpcitemtable{{\tt outputfile='filename'}}{%
{\tt outputfile='\ '} &:& For a blank character string (default) all output is 
written to standard output. \\  
{\tt outputfile='filename'} &:& Any other string usable as a file name redirects the output
to a file with the given filename.}%
\item[{\bf {\tt selprocess:}}] integer that selects the process to be calculated.
\cpcitemtable{{\tt selprocess=0}}{%
{\tt selprocess=0} &:& $\Pp \Pp \to \mathrm{j}\, \mathrm{j}\, \PH$ production (VBFH) (default),\\  
{\tt selprocess=1} &:& $\Pp \Pp \to \Pl^+  \Pnl  \PH$ production ($\PW^+\PH$),\\
{\tt selprocess=2} &:& $\Pp \Pp \to \Pnlbar \Pl^-  \PH$ production ($\PW^-\PH$), \\
{\tt selprocess=3} &:& $\Pp \Pp \to \Pl^+  \Pl^-  \PH$ production ($\PZ\PH$), \\
{\tt selprocess=4} &:& $\Pp \Pp \to \Pnlbar \Pnl  \PH$ production ($\PZ\PH$). \\
}%
A change of this flag sets all options and cuts to their respective default values. 
They can be changed thereafter.
In particular, for VBFH, the typical VBFH cuts are selected. To use \HAWK\ for Higgs strahlung
with hadronic vector-boson decay, one has to select {\tt selprocess=0} and change the respective
kinematic cuts appropriately thereafter.
\item[{\bf {\tt nevents:}}] integer that selects the number of generated weighted events.
\cpcitemtable{{\tt nevents=10000000}}{%
{\tt nevents=10000000} &:& is the default value.\\  
}%
The number of weighted events should be at least $10^7$ to guarantee
that the multi-channel integration yields reliable results and error estimates.
For histograms more events should be generated. For the published
distributions we used $10^9$ events.
\item[{\bf {\tt energy:}}] double-precision number that selects the centre-of-mass energy 
in the $\Pp\Pp/\Pp\Ppbar$ system in GeV. 
\cpcitemtable{{\tt energy=14000d0}}{%
{\tt energy=14000d0} &:& is the default value.\\  
}%
\item[{\bf {\tt sppbar:}}] integer that selects the hadron collider type.
\cpcitemtable{{\tt sppbar=0}}{%
{\tt sppbar=0} &:& $\Pp \Pp$ collider such as the LHC (default),\\  
{\tt sppbar=1} &:& $\Pp \Ppbar$ collider such as the Tevatron.\\
}%
For a $\Pp \Ppbar$ collider, the proton $\Pp$ is going in the $+z$ direction.
\end{cpcdescription}

\subsubsection{Input parameters}
\label{se:Input_parameters}
The following input parameters can be specified with double-precision values. Here, 
we also state the corresponding default values:
\cpctable{{\bf {\tt mmu=0.105658367d0 :}}}{%
{\bf {\tt gf=0.1166370d-04 }}&:& Fermi constant $G_\mathrm{F}$ in GeV$^{-2}$,\\
{\bf {\tt mz=91.1876d0 }}&:& Z-boson mass $\MZ$ in GeV,\\
{\bf {\tt gz=2.4952d0 }}&:& Z-boson width $\Gamma_\PZ$ in GeV,\\
{\bf {\tt mw=80.398d0 }}&:& W-boson mass $\MW$ in GeV,\\
{\bf {\tt gw=2.08872d0 }}&:& W-boson width $\Gamma_\PW$ in GeV,\\
{\bf {\tt mmu=0.105658367d0 }}&:& muon mass $m_\mu$ in GeV,\\
{\bf {\tt mt=172.5d0 }}&:& top-quark mass $\Mt$ in GeV,\\
{\bf {\tt mh=126.0d0 }}&:& Higgs-boson mass $\MH$ in GeV,\\
{\bf {\tt gh=4.21d-3 }}&:& Higgs-boson width $\Gamma_\PH$ in GeV,\\
{\bf {\tt sinthetac=0.225d0 }}&:& sine of the Cabibbo angle {$\theta_\mathrm{c}$}.
}%
For the weak gauge bosons $\PV=\PW,\PZ$,
the program reads the on-shell masses and widths and translates
them internally to the pole masses and widths,
\begin{eqnarray*}
M_V^{\mathrm{pole}} &=& M_{\PV}^{\mathrm{on-shell}}/
\sqrt{1+(\Gamma_{\PV}^{\mathrm{on-shell}}/M_{\PV}^{\mathrm{on-shell}})^2}\;,\\
\Gamma_{\PV}^{\mathrm{pole}} &=& \Gamma_{\PV}^{\mathrm{on-shell}}/
\sqrt{1+(\Gamma_{\PV}^{\mathrm{on-shell}}/M_{\PV}^{\mathrm{on-shell}})^2}\;.
\end{eqnarray*}
These values are then used
in propagators, the complex weak mixing angle, and other derived couplings.
The output provides the pole masses, not the input masses.

\noindent
The masses of the light fermions appear internally, but in the
  $\alpha_{G_\mathrm{F}}$ scheme used the results are practically 
independent of the specific values. Only for bare leptons in the final
state of the Higgs-strahlung process their masses
become relevant. Here, \HAWK\ uses the input value for the muon mass {\tt mmu},
as the muon case is the only one where the bare-lepton treatment is reasonable.

\noindent
The CKM matrix is always treated as real and block-diagonal, 
i.e.\ the first two generations mix with each other, but not with the third generation. 
In this approximation, the CKM matrix is fully specified by the Cabibbo angle.

\subsubsection{Recombination}
\label{se:Recombination}

IR-safe observables are constructed by properly defining jets in QCD. 
For VBFH, the recombination of quarks and gluons into jets is performed according 
to the $\kt$ algorithm~\cite{Blazey:2000qt}. The different variants of the employed
jet algorithm are discussed below (see {\tt ktpower}). Including EW corrections, also photons
are treated as partons in the jet algorithm for VBFH.
\begin{cpcdescription}
\item[{\bf {\tt ktpower:}}] integer that selects the type of $\kt$ algorithm.
\cpcitemtable{{\tt ktpower=-1}}{%
{\tt ktpower=1} &:& standard $\kt$ algorithm,\\  
{\tt ktpower=0} &:& Cambridge--Aachen algorithm,\\
{\tt ktpower=-1}&:& anti-$\kt$ algorithm.\\
}%
\item[{\bf {\tt dparameter=0.8d0 :}}] double-precision number that sets the 
$D$ parameter of the jet algorithm, i.e.\ the jet size (0.8d0 is the default value).
\item[{\bf {\tt etacut(parton)=5.0d0 :}}] double-precision number that specifies 
the pseu\-do-ra\-pi\-di\-ty cut on partons in the jet algorithm (5.0d0 is the default value).
Partons that do not pass the cut are not considered for recombination in the jet algorithm.
\end{cpcdescription}
For WH/ZH, there are no jets in the final state at LO,
and at NLO there is at most one jet or photon, so that no jet algorithm is applied here. 
However, the recombination of charged leptons and a bremsstrahlung photon is important 
if no perfect isolation of leptons can be achieved, as e.g.\ for electrons and
photons which appear as showers in the electromagnetic calorimeters of the detectors.
\begin{cpcdescription}
\item[{\bf {\tt sbarelep:}}] integer that steers lepton--photon isolation.
\cpcitemtable{{\tt sbarelep=1}}{%
{\tt sbarelep=1} &:& leptons are isolated from collinear photons (default),\\
{\tt sbarelep=0} &:& no such isolation is applied.\\
}%
\end{cpcdescription}
Technically, the lepton--photon isolation is performed according
to \citere{Dittmaier:2008md}.  As lepton--photon isolation makes
sense only in case of muons in the final state, the lepton mass in the
splittings is automatically set to the muon mass.  To investigate the
electron final state, {\tt sbarelep=0} is recommended. In this case,
leptons and photons are recombined:
\begin{cpcdescription}
\item[{\bf {\tt dgammaparameter=0.1d0 :}}] double-precision number that specifies the
$D$ parameter of lepton--photon recombination (0.1d0 is the default value). 
\end{cpcdescription}
For $R_{\gamma\Pl} < D$, 
the lepton and the photon are recombined. If this holds for more
than one lepton, the one with the smaller $R_{\gamma\Pl}$ is
recombined with the photon. The recombined particle,
which carries the total momentum of the two combined particles, 
is treated as a lepton. Here, $R_{\gamma\Pl}= 
\sqrt{(y_\Pl-y_\gamma)^2+(\varphi_\Pl-\varphi_\gamma)^2}$ is the usual separation in 
rapidity $y$ and angle $\varphi$ in the transverse plane.
               
For other recombination schemes the subroutines
{\tt vbfh\_recombination} in {\tt vbfh\_public.F} or
{\tt whzh\_recombination} in {\tt whzh\_public.F} have to be modified.

\subsubsection{Cuts}

The cuts on final-state jets or leptons can be set by a flag to predefined
values, but can also be chosen individually.
\begin{cpcdescription}
\item[{\bf {\tt scuts :}}] integer that governs the phase-space cuts.\\[1ex]
The meaning of {\tt scuts} depends on the chosen value for {\tt selprocess}. For
VBFH, i.e.\ {\tt selprocess=0}, the possible values correspond to:
\cpcitemtable{{\tt scuts=1}}{%
{\tt scuts=1} &:& standard cuts as defined in \citeres{Ciccolini:2007jr,Ciccolini:2007ec}
               (see also~\citere{Figy:2004pt}),
               jets are ordered according to $\pt$ of jets (default),\\
{\tt scuts=2} &:& standard cuts as defined in \citeres{Ciccolini:2007jr,Ciccolini:2007ec}
               (see also~\citere{Figy:2004pt}),
               jets are ordered according to the energy of jets,\\
{\tt scuts=0} &:& no cuts  (only useful for total cross sections).\\
}%
For Higgs strahlung, i.e.\ {\tt selprocess=1,2,3,4}, the possible values correspond to:
\cpcitemtable{{\tt scuts=1}}{%
{\tt scuts=1} &:& standard cuts for detected leptons 
as defined in \citere{Denner:2011id} (default),\\
{\tt scuts=2} &:& standard cuts for undetected leptons 
as defined in \citere{Denner:2011id},\\
{\tt scuts=0} &:& no cuts  (only useful for total cross section).\\
}%
\end{cpcdescription}
For VBFH, according to the choice for {\tt scuts}, in the following the leading jet $j_1$ is the
jet with the largest transverse momentum or largest energy. The next-to-leading
jet $j_2$ is the jet with the second largest transverse momentum or second largest energy, etc.

The above options set various cuts to their default values which are stated below.
They can be changed thereafter.
For other cut schemes the subroutines
         {\tt applycut} or {\tt applyhiggscut} in {\tt public.F} or
         {\tt applyjetcut} in {\tt vbfh\_public.F} or
         {\tt applylepcut} or {\tt applylephiggscut} in {\tt whzh\_public.F}
         have to be modified.    

All the standard cut parameters can be set/changed after choosing \linebreak ${\tt scuts}>0$. 
In the following the needed input is given together with the default values which
depend on {\tt scuts} and {\tt selprocess}. All the following cuts are 
specified by double-precision numbers, and dimensionful quantities are given in GeV. 
For VBFH, i.e.\ {\tt selprocess=0}, the standard cuts are as follows:
\cpctable{{\bf {\tt dphicut(jet1,higgs)=0d0 }}}{
{\bf {\tt ptcut(jet1)=20d0 }}&:& lower $\pt$ cut on $j_1$ (leading jet),\\
{\bf {\tt ptcut(jet2)=20d0 }}&:& lower $\pt$ cut on $j_2$ (subleading jet),\\
{\bf {\tt ycut(jet1)=4.5d0 }}&:& upper $|y|$ cut on $j_1$,\\
{\bf {\tt ycut(jet2)=4.5d0 }}&:& upper $|y|$ cut on $j_2$,\\
{\bf {\tt dycut(jet,jet)=4.0d0 }}&:& lower $\Delta y$ cut between $j_1$ and $j_2$,\\
{\bf {\tt ptmax(jet1)=1d30 }}&:& upper $\pt$ cut on $j_1$,\\
{\bf {\tt ptmax(jet2)=1d30 }}&:& upper $\pt$ cut on $j_2$,\\
{\bf {\tt ptmax(jet3)=1d30 }}&:& upper $\pt$ cut on $j_3$,\\
{\bf {\tt ymin(jet1)=0d0 }}&:& lower $|y|$ cut on $j_1$,\\
{\bf {\tt ymin(jet2)=0d0 }}&:& lower $|y|$ cut on $j_2$,\\
{\bf {\tt ymin(jet3)=0d0 }}&:& lower $|y|$ cut on $j_3$,\\
{\bf {\tt ptcut(visible)=0d0 }}&:&  lower cut on $|\mathbf{p}_\mathrm{T,j_1}+\mathbf{p}_{\mathrm{T,j_2}}+\mathbf{p}_{\mathrm{T,\PH}}|$ (vector sum),\\
{\bf {\tt ptmax(visible)=1d30 }}&:& upper cut on $|\mathbf{p}_\mathrm{T,j_1}+\mathbf{p}_{\mathrm{T,j_2}}+\mathbf{p}_{\mathrm{T,\PH}}|$ (vector sum),\\
{\bf {\tt mlcut(jet,jet)=0d0 }}&:&  lower cut on invariant mass of $j_1$ and $j_2$,\\
{\bf {\tt mucut(jet,jet)=1d30 }}&:& upper cut on invariant mass of $j_1$ and $j_2$,\\
{\bf {\tt dphicut(jet1,higgs)=0d0 }}&:& lower cut on azimuthal angle between $j_1$ and the Higgs boson,\\
{\bf {\tt dphicut(jet2,higgs)=0d0 }}&:& lower cut on azimuthal angle between $j_2$ and the Higgs boson,\\
{\bf {\tt ptcut(higgs)=0d0 }}&:&  lower $\pt$ cut on the Higgs boson,\\
{\bf {\tt ycut(higgs)=1d10 }}&:&  upper $|y|$ cut on the Higgs boson,\\
{\bf {\tt ecut(higgs)=0d0 }}&:&   lower energy cut on the Higgs boson,\\
{\bf {\tt mlcut(higgs)=1d-2 }}&:& lower cut on the invariant mass of an off-shell Higgs boson,\\
{\bf {\tt mucut(higgs)=1d30 }}&:& upper cut on the invariant mass of an off-shell Higgs boson.
}
In the above, $y$ denotes the rapidity of the specified final-state object and $\pt$ 
its transverse momentum. In particular, the lower cut on the rapidity difference of the 
leading jets, {\tt dycut(jet,jet)=4.0d0}, is specific to
select the VBFH process. There is one more VBFH-specific cut:
\begin{cpcdescription}
\item[{\bf {\tt hemispherecut : }}] integer that governs the condition
$y(j_1) y(j_2) < 0$.
\cpcitemtable{{\tt hemispherecut=1}}{%
{\tt hemispherecut=1} &:& $y(j_1) y(j_2) < 0$ is required for 
$j_1$ and $j_2$ (default),\\
{\tt hemispherecut=0} &:& $y(j_1) y(j_2) < 0$ cut is not applied.\\
}%
\end{cpcdescription}
{To implement a jet veto,} {\tt ptmax} and {\tt ymin} can be used in combination to define 
final states without detectable jets, 
i.e.\ a jet is considered undetectable if either $\pt(\mathrm{jet}) > {\tt ptmax(jet)}$ 
or $|y(\mathrm{jet})| < {\tt ymin(jet)}$ fails.
The cuts for $j_1$ apply to all jets, those for $j_2$ for all but the leading
jet, those for $j_3$ {only veto a third jet}.
Set {\tt ptmax(jet1)=1d30} and/or {\tt ymin(jet1)=0d0} to switch off these cuts,
and likewise for $j_2$ and $j_3$.

For WH/ZH, i.e.\ {\tt selprocess=1,2,3,4}, the standard cuts partly depend on the value
of {\tt scuts}. For {\tt scuts=1}, they are given by:
\cpctable{{\bf {\tt ptcut(lep)=20d0 }}}{
{\bf {\tt ptcut(lep)=20d0 }}&:& lower $\pt$ cut on leptons,\\
{\bf {\tt ycut(lep)=2.5d0 }}&:& upper $|y|$ cut on leptons,\\
{\bf {\tt ptmax(lep)=1d10 }}&:& upper $\pt$ cut on leptons,\\
{\bf {\tt ymin(lep)=0d0 }}&:&   lower $|y|$ cut on leptons.
}%
For {\tt scuts=2}, they are given by:
\cpctable{{\bf {\tt ptmax(lep)=20d0 }}}{
{\bf {\tt ptcut(lep)=0d0 }}&:&  lower $\pt$ cut on leptons,\\
{\bf {\tt ycut(lep)=1d10 }}&:&  upper $|y|$ cut on leptons,\\
{\bf {\tt ptmax(lep)=20d0 }}&:& upper $\pt$ cut on leptons,\\
{\bf {\tt ymin(lep)=2.5d0 }}&:& lower $|y|$ cut on leptons.
}%
The default cuts, which are the same for {\tt scuts=1} and {\tt scuts=2}, are given by:
\cpctable{{\bf {\tt mucut(lep,lep)=1d10 }}}{
{\bf {\tt ptcut(V)=190d0 }}&:&      lower $\pt$ cut on $\PV$,\\
{\bf {\tt ptcut(miss)=25d0 }}&:&    lower cut on missing-$\pt$ ($\pt$ of neutrino(s)),\\
{\bf {\tt drcut(lep,lep)=0d0 }}&:&  lower cut on $R_{\Pl\Pl}$ between leptons,\\
{\bf {\tt drcut(jet,lep)=0d0 }}&:&  lower cut on $R_{\Pl j}$ between lepton and jet,\\
{\bf {\tt mlcut(lep,lep)=0d0 }}&:&  lower cut on invariant mass between leptons,\\
{\bf {\tt mucut(lep,lep)=1d10 }}&:& upper cut on invariant mass between leptons,\\
{\bf {\tt ecut(lep)=0d0 }}&:&       lower energy cut on leptons,\\
{\bf {\tt ptcut(higgs)=200d0 }}&:&  lower $\pt$ cut on the Higgs boson,\\
{\bf {\tt ycut(higgs)=1d10 }}&:&    upper $|y|$ cut on the Higgs boson,\\
{\bf {\tt ecut(higgs)=0d0 }}&:&     lower energy cut on the Higgs boson.
}
{In analogy to a jet veto,} {\tt ptmax} and {\tt ymin} are used in combination 
to define final states without detectable leptons, 
i.e.\ a lepton is considered undetectable if either $\pt(\Pl) > {\tt ptmax(lep)}$ 
or $|y(\Pl)| < {\tt ymin(lep)}$ fails. Set ${\tt ptmax(lep)=1d10}$ 
and/or ${\tt ymin(lep)=0d0}$ to switch off this cut. {These cuts are 
particularly useful to calculate the contribution of the
$\Pp\Pp\to \PH \Pl^+\Pnl$ and
$\Pp\Pp\to \PH \Pnlbar \Pl^-$ processes to the 
$\PH \Pnlbar\Pnl$ final state, i.e.\ a Higgs boson plus missing $\pt$, due to a missed charged lepton.}

\subsubsection{Off-shell Higgs boson, Higgs decays and corresponding cuts}
\label{se:off-shell}

Within \HAWK, there are several options that allow to choose between on-shell and off-shell 
Higgs-boson production and to specify the details of the off-shell treatment.
\begin{cpcdescription}
\item[{\bf {\tt shtr :}}] integer that selects on-shell or off-shell Higgs production.
\cpcitemtable{{\tt shtr=0}}{%
{\tt shtr=0} &:& on-shell Higgs boson (default),\\
{\tt shtr=1} &:& off-shell Higgs boson,\\
{\tt shtr=2} &:& off-shell Higgs boson decaying into a pair of massless singlets
(for VBFH, i.e.\ {\tt selprocess=0}, only).
}%
\end{cpcdescription}
For {\tt shtr=1} the external Higgs boson is off shell, and the
resonance is treated according to the flag {\tt shbw} (see below).
For {\tt shtr=2} additionally an isotropic Higgs decay into a pair of
massless scalar singlets is included. This can be used to mimic any two-body
decay into massless particles, e.g.\ the decay into photons.
For {\tt shtr=1} or {\tt shtr=2} the width of the Higgs boson has to be given as input,
or the width can be calculated internally (see {\tt gh} and {\tt sgh} below).
For {\tt shtr=2}, in addition, the branching ratio for the decay into singlets has to be specified. 
For an off-shell Higgs boson the EW corrections are
calculated after an on-shell projection of the momenta
(required by gauge invariance).

The following flags dictate the details of the off-shell treatment:
\begin{cpcdescription}
\item[{\bf {\tt shbw :}}] integer that determines the treatment of the off-shell Higgs 
propagator for {\tt shtr=1} or {\tt shtr=2}.
\cpcitemtable{{\tt shbw=1}}{%
{\tt shbw=0} &:& standard Breit--Wigner with mass {\tt mh} and (constant) width {\tt gh},\\
{\tt shbw=1} &:& off-shell propagator according to \citere{Anastasiou:2011pi}, Eq.~(4.6) 
(default if {\tt shtr}${}\ne0$). 
Note that choosing the flag {\tt sgh=0} is not allowed in this case.
}%
\end{cpcdescription}
\begin{cpcdescription}
\item[{\bf {\tt sgh :}}] integer that determines the treatment of the Higgs-boson width.
\cpcitemtable{{\tt sgh=0}}{%
{\tt sgh=0} &:& use Higgs-boson width as specified in input by {\tt gh},\\
{\tt sgh=1} &:& {\tt gh} is set for the input Higgs mass using an interpolation
                of the results in \citere{Dittmaier:2011ti},
                (based on a routine of G. Passarino), the input value {\tt gh} is not used,\\
{\tt sgh=2} &:& {\tt gh} is calculated according to the complex-pole scheme~\cite{Dittmaier:2011ti} 
                (using cpHTO11.f by G. Passarino) (default), the input value {\tt gh} is not used.               
}%
\end{cpcdescription}
\begin{cpcdescription}
\item[{\bf {\tt gh=0.00421d0 :}}] double-precision number for the Higgs-boson width (default for 
$\MH=126\GeV$).
\end{cpcdescription}
\begin{cpcdescription}
\item[{\bf {\tt Hbr=1d0 :}}] double-precision number for the branching ratio for Higgs decay into singlets. 
{\tt Hbr} simply rescales the complete cross section.
\end{cpcdescription}
If the Higgs decay into singlets is included, the following cuts on
the decay products $d_{1,2}$ of the Higgs boson are available: 
\cpctable{{\bf {\tt ptcut(decp)=0d0 :}}}{
{\bf {\tt ptcut(decp)=0d0}}&:& lower
$\pt$ cut on Higgs-decay products $d_{1,2}$,\\
{\bf {\tt ycut(decp)=1d10}}&:& upper
$|y|$ cut on Higgs-decay products $d_{1,2}$.%
}%
\begin{cpcdescription}
\item[{\bf {\tt higgsbetweenjets :}}] integer that dictates whether the event is cut if Higgs-decay 
products $d_{1,2}$ are not produced in between the tagging jets with respect to rapidity.
\cpcitemtable{{\tt higgsbetweenjets=0}}{%
{\tt higgsbetweenjets=0} &:& the selection criterion
$\min(y(j_1),y(j_2))<y(d_{1,2})<\max(y(j_1),y(j_2))$
 is not applied,\\
{\tt higgsbetweenjets=1} &:& the selection criterion 
$\min(y(j_1),y(j_2))<y(d_{1,2})<\max(y(j_1),y(j_2))$ is applied.
}%
\end{cpcdescription}

\subsubsection{Parton distribution functions}
\label{se:PDF}

\HAWK\ supports the LHAPDF interface to evaluate parton distribution functions (PDFs) for 
hadronic cross sections. In addition, a few built-in PDF sets are available:
\begin{cpcdescription}
\item[{\bf {\tt spdf :}}] integer that determines the usage of PDFs.
\cpcitemtable{{\tt spdf=99 }}{%
{\tt spdf=1} &:& MRST2004 parton distributions are used, \\
{\tt spdf=11} &:& CTEQ6M parton distributions are used, \\
{\tt spdf=14} &:& CTEQ6L1 parton distributions are used, \\
{\tt spdf=99} &:& the LHAPDF interface is used, i.e.\ the PDF set is selected
via {\tt pdfname} or {\tt pdfpath} as described in the following.
}%
\end{cpcdescription}
A photon PDF~\cite{Martin:2004dh,Carrazza:2013bra,Spiesberger:1994dm}, 
which is needed for the cross-section calculation of partonic 
processes with photons in the initial state, is only available for 
{\tt spdf=1} (using the MRST2004QED set~\cite{Martin:2004dh}) or for {\tt spdf=99} 
if appropriate parton distributions are used. This is the case for the default 
PDF set NNPDF2.3QED~\cite{Carrazza:2013bra}.
For other PDF sets only quark and gluon PDFs are taken into account. In this case, the 
prediction for the photon-initiated processes is trivially zero.

To select a PDF set using the LHAPDF interface, there are the following 
input options:
\begin{cpcdescription}
\item[{\bf {\tt pdfname=NNPDF23\_nlo\_as\_0118\_qed.LHgrid :}}] a string that specifies the \linebreak
name of the PDF set to be used. The corresponding file has to be in
the directory {\tt PDFsets} of the
PDF installation. The path to the directory {\tt PDFsets} has to be set in advance,
e.g.\ by {\tt export LHAPATH=/...\linebreak/share/lhapdf/PDFsets} in \tt .bashrc.
\end{cpcdescription}
\begin{cpcdescription}
\item[{\bf {\tt pdfpath='/.../PDFsets/NNPDF23\_nlo\_as\_0118\_qed.LHgrid' :}}] 
a string \linebreak that specifies the name of the the PDF set including the full path to the file. 
Note the apostrophes around the specified {\tt pdfpath}.
\end{cpcdescription}
\begin{cpcdescription}
\item[{\bf {\tt pdfmember :}}] an integer that specifies the member of the selected family of PDF sets.
Usually, {\tt pdfmember=0} (default) is the central value PDF set. The allowed range of {\tt pdfmember}
depends on the selected PDF set and is provided by LHAPDF.
\end{cpcdescription}
For PDF uncertainty calculations one can simultaneously calculate
additional cross sections for more than one PDF set if one uses LHAPDF
with a  {\tt *.LHgrid} set (40 sets will only double the runtime).
The result for the cross section (including all the corrections) will
be written to the specified output together with the results for the
central PDF member. Distributions are only produced for the central member
specified by {\tt pdfmember}. The range of PDF members is set by the following input:
\begin{cpcdescription}
\item[{\bf {\tt pdfmemberfrom :}}] integer that selects the beginning of the range of
evaluated PDF members.
\end{cpcdescription}
\begin{cpcdescription}
\item[{\bf {\tt pdfmemberto :}}] integer that selects the end of the range of
evaluated PDF members.
\end{cpcdescription}
The default is {\tt pdfmemberfrom=1} and {\tt pdfmemberto=0} which means that only 
the set selected via {\tt pdfmember} is evaluated, no additional sets are used.
If only a {\tt *.LHpdf} implementation is available one should not use
the simultaneous calculation with different PDF sets but evaluate
the different sets sequentially, otherwise the runtime explodes due to
repeated PDF initializations.

\subsubsection{Factorization and renormalization scales}

The QCD factorization and renormalization scales and schemes are set by the following input:
\begin{cpcdescription}
\item[{\bf {\tt sfactqcd :}}] integer that selects the QCD factorization scheme.
\cpcitemtable{{\tt sfactqcd=1}}{%
{\tt sfactqcd=1} &:& $\overline{\mathrm{MS}}$ factorization is used with respect to QCD (default), \\
{\tt sfactqcd=2} &:& DIS factorization is used with respect to QCD.
}%
\end{cpcdescription}
\begin{cpcdescription}
\item[{\bf {\tt sfactqed :}}] integer that selects the QED factorization scheme.
\cpcitemtable{{\tt sfactqed=1}}{%
{\tt sfactqed=1} &:& $\overline{\mathrm{MS}}$ factorization is used with respect to QED, \\
{\tt sfactqed=2} &:& DIS factorization is used with respect to QED (default).
}%
\end{cpcdescription}
\begin{cpcdescription}
\item[{\bf {\tt qcdfacscalefac=1d0 :}}] double-precision number that multiplies the default 
QCD factorization scale.
\item[{\bf {\tt qcdrenscalefac=1d0 :}}] double-precision number that multiplies the default 
QCD renormalization scale.
\item[{\bf {\tt qedfacscalefac=1d0 :}}] double-precision number that multiplies the default 
QED factorization scale.
\end{cpcdescription}
The last three parameters rescale the factorization and renormalization scales of QCD
and the factorization scale of QED which are set by default to:
\[
\begin{array}{lll}
$\MW$  &   \mathrm{for\,\, VBFH\,\, production} &({\tt selprocess=0}),\\
$\MW+\MH$ & \mathrm{for\,\, WH\,\, production}  &({\tt selprocess=1,2}),\\
$\MZ+\MH$ & \mathrm{for\,\, ZH\,\, production}  &({\tt selprocess=3,4}).
\end{array}
\]
Phase-space-dependent (often called dynamical) scale choices are not supported by the current version of \HAWK.

\subsubsection{Switches for different LO/NLO contributions}

The following switches can be used to switch on or off 
the LO contributions and various
contributions to the NLO corrections.
By default all contributions are taken into account.
\begin{cpcdescription}
\item[{\bf {\tt sborn :}}] integer that includes/excludes the Born contribution. 
\cpcitemtable{{\tt sborn=1}}{%
{\tt sborn=0} &:& tree-level contributions not included,\\
{\tt sborn=1} &:& tree-level contributions included (default).
}%
\end{cpcdescription}
\begin{cpcdescription}
\item[{\bf {\tt sbini :}}] integer that includes/excludes contributions from initial-state $\Pb$~quarks.%
\cpcitemtable{{\tt sbini=1}}{%
{\tt sbini=0} &:& initial-state $\Pb$-quark contributions not included,\\
{\tt sbini=1} &:& initial-state $\Pb$-quark contributions included (default).}%
\end{cpcdescription}
\begin{cpcdescription}
\item[{\bf {\tt sbfin :}}] integer that includes/excludes contributions from final-state $\Pb$~quarks.%
\cpcitemtable{{\tt sbfin=1}}{%
{\tt sbfin=0} &:& final-state $\Pb$-quark contributions not included,\\
{\tt sbfin=1} &:& final-state $\Pb$-quark contributions included (default).}%
\end{cpcdescription}
The contributions included/excluded via {\tt sbini} and {\tt sbfin} are only calculated at LO,
since they contribute only at the per-cent level.
\begin{cpcdescription}
\item[{\bf {\tt sew :}}] integer that includes/excludes the EW corrections.%
\cpcitemtable{{\tt sew=1}}{%
{\tt sew=0} &:& EW corrections not included,\\
{\tt sew=1} &:& EW corrections included (default).}%
\end{cpcdescription}
\begin{cpcdescription}
\item[{\bf {\tt spinc :}}] integer that includes/excludes incoming photon contributions.
\cpcitemtable{{\tt spinc=1}}{%
{\tt spinc=0} &:& incoming photon contributions not included,\\
{\tt spinc=1} &:& incoming photon contributions included (default), if the chosen PDF set
includes a photon PDF (see \refse{se:PDF}).}%
\end{cpcdescription}
\begin{cpcdescription}
\item[{\bf {\tt sqcd :}}] integer that includes/excludes the QCD corrections.
\cpcitemtable{{\tt sqcd=1}}{%
{\tt sqcd=0} &:& QCD corrections not included,\\
{\tt sqcd=1} &:& QCD corrections included (default).}%
\end{cpcdescription}
For Higgs production in vector-boson fusion ({\tt selprocess=0}), the
additional switches described in the rest of this section are available 
for contributions to NLO corrections:
\begin{cpcdescription}
\item[{\bf {\tt shh2 :}}] integer that includes/excludes 2-loop heavy-Higgs corrections.%
\cpcitemtable{{\tt shh2=1}}{%
{\tt shh2=0} &:& 2-loop heavy-Higgs corrections not included,\\
{\tt shh2=1} &:& 2-loop heavy-Higgs corrections included (default). 
These contributions are taken from~\citere{Frink:1996sv}.}%
\end{cpcdescription}
\begin{cpcdescription}
\item[{\bf {\tt sqcddiag :}}] integer that includes/excludes diagonal QCD corrections.
\cpcitemtable{{\tt sqcddiag=1}}{%
{\tt sqcddiag=0} &:& diagonal QCD corrections not included,\\
{\tt sqcddiag=1} &:& diagonal QCD corrections included (default).}%
\end{cpcdescription}
\begin{cpcdescription}
\item[{\bf {\tt sqcdnondiag :}}] integer that includes/excludes non-diagonal QCD corrections.
\cpcitemtable{{\tt sqcdnondiag=1}}{%
{\tt sqcdnondiag=0} &:& non-diagonal QCD corrections not included,\\
{\tt sqcdnondiag=1} &:& non-diagonal QCD corrections included (default).}%
\end{cpcdescription}
\begin{cpcdescription}
\item[{\bf {\tt sqcdggfus :}}] integer that includes/excludes interference with gluon-fusion contributions.%
\cpcitemtable{{\tt sqcdggfus=1}}{%
{\tt sqcdggfus=0} &:& interference with gluon-fusion contributions not included,\\
{\tt sqcdggfus=1} &:& interference with gluon-fusion contributions included (default).
}%
\end{cpcdescription}
\begin{cpcdescription}
\item[{\bf {\tt sqcdgsplit :}}] integer that includes/excludes interference with gluon-splitting contributions.
\cpcitemtable{{\tt sqcdgsplit=1}}{%
{\tt sqcdgsplit=0} &:& interference with gluon-splitting contributions not included,\\
{\tt sqcdgsplit=1} &:& interference with gluon-splitting contributions included (default).
}%
\end{cpcdescription}
The precise definition of these contributions can be found in \citere{Ciccolini:2007ec}.
If {\tt sch2=0} (see below), the diagonal QCD corrections are always excluded,
i.e.\ {\tt sqcddiag} has no effect.
If {\tt schint=0} (see below), the non-diagonal QCD corrections, the interferences
with gluon fusion, and gluon splitting are always excluded,
i.e.\ {\tt sqcdnondiag}, {\tt sqcdggfus}, and {\tt sqcdgsplit} have no effect.
If {\tt shvv>0} (see below), {\tt sqcdnondiag}, {\tt sqcdggfus}, {\tt sqcdgsplit} are set to zero,
i.e.\ these contributions are not supported for anomalous Higgs couplings.

The following switches for VBFH can be used to switch on or off
contributions connected to $s$- or $t$-channel diagrams,
squared diagrams, or interferences. By default all contributions 
are included. Since the exclusion of diagrams is not a physical 
procedure, you should only use this option if you know what you are doing. 
\begin{cpcdescription}
\item[{\bf {\tt sscha :}}] integer that includes/excludes the $s$-channel contribution.%
\cpcitemtable{{\tt sscha=1}}{%
{\tt sscha=0} &:& $s$-channel contribution not included,\\
{\tt sscha=1} &:& $s$-channel contribution included (default).}%
\end{cpcdescription}
\begin{cpcdescription}
\item[{\bf {\tt stcha :}}] integer that includes/excludes the
  $t$-channel and $u$-channel contributions.%
\cpcitemtable{{\tt stcha=1}}{%
{\tt stcha=0} &:& $t/u$-channel contributions not included,\\
{\tt stcha=1} &:& $t/u$-channel contributions included (default).}%
\end{cpcdescription}
\begin{cpcdescription}
\item[{\bf {\tt sch2 :}}] integer that includes/excludes contributions due to 
squared LO diagrams and the corresponding corrections (defined by the fermion-number flow).%
\cpcitemtable{{\tt sch2=1}}{%
{\tt sch2=0} &:& contributions due to squared diagrams not included,\\
{\tt sch2=1} &:& contributions due to squared diagrams included (default).}%
\end{cpcdescription}
\begin{cpcdescription}
\item[{\bf {\tt schint :}}] integer which includes/excludes contributions due to interferences
(defined as complement to the squared LO diagrams and their corrections steered by {\tt sch2}).%
\cpcitemtable{{\tt schint=1}}{%
{\tt schint=0} &:& contributions due to interferences not included,\\
{\tt schint=1} &:& contributions due to interferences included (default).
}%
\end{cpcdescription}
To take only squared $t$-channel (and $u$-channel) diagrams into account, which corresponds
to the definition of the total VBFH cross section in 
\citeres{Dittmaier:2011ti,Dittmaier:2012vm,Heinemeyer:2013tqa} use
{\tt sscha=0}, {\tt stcha=1}, {\tt sch2=1}, and {\tt schint=0}.
QCD contributions are in addition steered by the switches {\tt sqcddiag},
{\tt sqcdnondiag}, {\tt sqcdggfus}, and {\tt sqcdgsplit} discussed above.

\subsubsection{Anomalous HVV couplings}

\HAWK\ supports predictions with anomalous $\PH\PV\PV$ couplings for
$\PV=\PW,\PZ$. Our implementation of anomalous couplings follows the (modified) parameterization of
\citere{Hankele:2006ma}. In addition the 
Standard Model $\PH\PV\PV$ coupling can be rescaled via {\tt rsm}. In this parameterization, 
the different (anomalous) couplings are related to the parameters {\tt d}, {\tt db},
{\tt dt}, and {\tt dtb} via
\begin{longtable}{lll}
{\tt a1hww} &$= {\tt rsm} \, \MW/\sw$                  &$= \mathrm{SM\,\, HWW\,\, coupling}$,\\
{\tt a2hww} &$= 2\,{\tt d}\, /(\sw\MW)$                &$= 2\,g^{(2)}_{\PH\PW\PW}$,\\
{\tt a3hww} &$= 2\,{\tt dt}/(\sw\MW)$                  &$= 2\,\tilde{g}^{(2)}_{\PH\PW\PW}$,\\
{\tt a1haa} &$= 0$                             &$= \PH\gamma\gamma \,\, \mathrm{like}\,\, \PH\PZ\PZ\,\, \mathrm{in}\,\, \mathrm{SM}$,\\
{\tt a2haa} &$= 4\,({\tt d}\, \sw^2+{\tt db}\, \cw^2)/(2\sw\MW)$ &$= 4\, g_{\PH\gamma\gamma}$,\\
{\tt a3haa} &$= 4\,({\tt dt}\,\sw^2+{\tt dtb}\,\cw^2)/(2\sw\MW)$ &$= 4\, \tilde{g}_{\PH\gamma\gamma}$,\\
{\tt a1haz} &$= 0$                             &$= \PH\PZ\gamma \,\,\mathrm{like}\,\, \PH\PZ\PZ \,\,\mathrm{in}\,\, \mathrm{SM}$,\\
{\tt a2haz} &$= -2\cw({\tt d} -{\tt db} )/\MW$         &$= 2 g^{(2)}_{\PH\PZ\gamma}$,\\
{\tt a3haz} &$= -2\cw({\tt dt}-{\tt dtb})/\MW$         &$= 2\tilde{g}^{(2)}_{\PH\PZ\gamma}$,\\
{\tt a1hzz} &$= {\tt rsm}\, \MW/(\sw\cw^2)$                    &$= \mathrm{SM}\,\, \PH\PZ\PZ\,\, \mathrm{coupling}$,\\
{\tt a2hzz} &$= 4({\tt d}\, \cw^2+{\tt db}\, \sw^2)/(2\sw\MW)$ &$= 4\,g^{(2)}_{\PH\PZ\PZ}$,\\
{\tt a3hzz} &$= 4({\tt dt}\,\cw^2+{\tt dtb}\,\sw^2)/(2\sw\MW)$ &$= 4\,\tilde{g}_{\PH\PZ\PZ}$,
\end{longtable}
\noindent
where the notation for the coupling constants corresponds to the Feynman rule
\[
\mathrm{i}\,{\tt a1hvv}\,g_{\mu\nu}
+ \mathrm{i}\,{\tt a2hvv}\,(-k_1\cdot k_2g_{\mu\nu}+k_{1\nu}k_{2\mu})
+ \mathrm{i}\,{\tt a3hvv}\,\epsilon_{\rho\sigma\mu\nu}k_1^\rho k_2^\sigma
\]
for $\PH\PV_1(k_{1\mu})\PV_2(k_{2\nu})$ and the
$g_{\PH\PV\PV}$'s correspond to the notation in \citere{Hankele:2006ma}. Note the sign changes 
in {\tt a2haz} and {\tt a3haz} due to our conventions of SM couplings
which follow \citeres{Bohm:1986rj,Denner:1991kt}. Cosine and sine of
the electroweak mixing angle are defined as $\cw=\MW/\MZ$ and
$\sw=\sqrt{1-\cw^2}$, respectively.
The input for the anomalous couplings is governed by the following switch:
\begin{cpcdescription}
\item[{\bf {\tt shvv :}}] integer that enables/disables anomalous $\PH\PV\PV$ couplings.
\cpcitemtable{{\tt shvv=1}}{%
{\tt shvv=0} &:& anomalous $\PH\PV\PV$ couplings disabled (default),\\
{\tt shvv=1} &:& anomalous $\PH\PV\PV$ couplings enabled, input for parameters 
{\tt d}, {\tt db}, {\tt dt}, {\tt dtb} expected,\\
{\tt shvv=2} &:& anomalous $\PH\PV\PV$ couplings enabled, direct input for {\tt a1hww}, $\ldots$, {\tt a3hzz} 
expected.
}%
\end{cpcdescription}
According to {\tt shvv}, one either specifies {\tt d}, {\tt db}, etc.\ or 
{\tt a1hww}, {\tt a2hww}, etc. as double-precision numbers. By default, all 
anomalous couplings are set to zero. One can also specify {\tt rsm} which by default 
is set to one.

The anomalous couplings to the neutral gauge bosons are switched
off for small momentum transfer with a form factor
$|s_1| |s_2| / ( m_0^2 + |s_1| )/ ( m_0^2 + |s_2| )$
to avoid IR singularities from anomalous couplings, where
$s_1$ and $s_2$ are the virtualities of the two intermediate $\PW$ and
$\PZ$~bosons and $m_0=1 \GeV$ is used.

If ${\tt shvv}>0$, {\tt sqcdnondiag}, {\tt sqcdggfus}, and {\tt sqcdgsplit} 
are set to zero. These QCD corrections 
(which are tiny in the SM) are not supported for anomalous
Higgs couplings.

To control the scale dependence of the anomalous couplings, there is the option to use a form factor:
\begin{cpcdescription}
\item[{\bf {\tt lambdahvv=-2d0 :}}] double-precision number that sets 
the mass scale in the form factor for anomalous couplings
in GeV.
\cpcitemtable{{\tt $lambdahvv>0$}}{%
{\tt ${\tt lambdahvv}>0$} &:&
form factor= \hfill \mbox{}\linebreak${\tt lambdahvv^4}/
({\tt lambdahvv^2} \!+\! |s_1| )/ ({\tt lambdahvv^2} \!+\! |s_2| )$,
\\
{\tt ${\tt lambdahvv}<0$} &:& form factor=1   (formal limit {\tt lambdahvv} $\to \infty$).
}%
\end{cpcdescription}

\subsubsection{Technical parameters for the Monte Carlo integration}

In order to produce statistically independent \HAWK\ results, one has to use
independent random-number sets which can be obtained with the following input:
\begin{cpcdescription}
\item[{\bf {\tt ranluxseed=0 :}}] integer that yields different random number seeds
by choosing different positive integers.
\end{cpcdescription}

\subsubsection{Parameters steering the output}

The level of detail of the output generated by \HAWK\ is steered by the following switches:
\begin{cpcdescription}
\item[{\bf {\tt lnoutmc :}}] integer that determines the level of detail of the 
\HAWK\ output.%
\cpcitemtable{{\tt lnoutmc=1}}{%
{\tt lnoutmc=1} &:& standard output of Monte Carlo, \\
{\tt lnoutmc=3} &:& intermediate output of Monte Carlo, \\
{\tt lnoutmc=5} &:& full output of Monte Carlo.
}%
\end{cpcdescription}
\begin{cpcdescription}
\item[{\bf {\tt shisto :}}] integer that determines the generation of
  histograms for differential distributions.
\cpcitemtable{{\tt shisto=1}}{%
{\tt shisto=0} &:& no histograms are produced, \\
{\tt shisto=1} &:& histograms are produced (default).
}%
\end{cpcdescription}
By default, histograms for standard hadron-collider observables relevant for
the selected Higgs-production process are generated. 
Additional histograms can be implemented by adapting the
subroutines {\tt settings\_Hjj} in \linebreak{\tt vbfh\_public.f} or
the subroutines {\tt settings\_Hll}, {\tt settings\_Hlv}, \linebreak {\tt settings\_Hvl}, or
{\tt settings\_Hvv} in {\tt whzh\_public.f} depending on \linebreak {\tt selprocess}.

\subsection{Parallel execution using the MPI standard}
\label{se:mpi}

\HAWK\ supports parallel execution using MPI. To use the parallel
version set {\tt FC=\$(MPIFC)} in the {\tt Makefile}, where {\tt MPIFC} should be your MPI
Fortran compiler. The parallel version of \HAWK\ has been tested using
the Intel Fortran compiler 11.1 and SUN's MPI 8.2.
Using \HAWK\ with MPI, the input cannot be provided via standard
input ({\tt ./hawk-2.0 < inputfile}), but has to be put into a file with the
literal name {\tt inputfile} in the directory of the {\tt hawk-2.0} executable.

\section{\HAWK\ output and sample runs}
\label{se:output}

By default, all output is written to standard output.
An {\tt outputfile} can, however, be specified in the inputfile (see \refse{se:input}).

The output includes the process that has been selected via {\tt selprocess} and the SM
parameters which have been used. The values for anomalous couplings and information 
on the treatment of an off-shell Higgs boson are provided if relevant (see \refse{se:input}). 
In addition, information about the Monte Carlo integration is 
provided, including the number of events generated during the run, collider information, the PDF choice, 
scale choices, included contributions, calculated corrections, the jet algorithm, and applied cuts.
Information on the seed for the random numbers is given as well.

\HAWK\ provides intermediate results for monitoring the progress of the Monte Carlo integration. 
The central final result for the total cross section can be found under
\cpctable{{\tt incoming photon cross section:}}{
{\tt Summary of results}:&& \\
{\tt Born cross section:}&&   LO\\
{\tt complete cross section:}&&   LO + NLO EW + NLO QCD\\
{\tt QCD corr.\ to cross section:}&&   NLO QCD = virtual QCD + real QCD + incoming gluon\\
{\tt incoming gluon cross section:}&&   incoming gluon\\
{\tt virt+real QCD cross section:}&&   virtual QCD + real QCD\\
{\tt virt.\ QCD cross section:}&&   virtual QCD\\
{\tt outgoing gluon cross section:}&&   real QCD\\
{\tt ew corr.\ to cross section:}&&   NLO EW  = virtual EW  + real EW + incoming photon\\
{\tt incoming photon cross section:}&&  incoming photon\\
{\tt virt+real ew cross section:}&&   virtual EW  + real EW\\
{\tt virt.\ ew cross section:}&&   virtual EW\\
{\tt outgoing photon cross section:}&&  real EW
}
All results for cross sections are given in femtobarns (fb). 
In addition, the relative corrections, normalized to the LO cross section, are provided.

\subsection{Differential distributions with \HAWK}
\label{se:Differential}

For {\tt shisto=1}, the program produces a set of histograms for
differential distributions of standard hadron-collider observables. The produced
histograms are listed at the end of the \HAWK\ output together with 
the range of the binned differential observable. The distributions which are available 
independently of the selected process are written to the files
\cpctable{{\tt dat.etah}}{
{\tt dat.pth}       &&    distribution in transverse momentum of Higgs,\\
{\tt dat.yh}        &&    distribution in rapidity of Higgs,\\
{\tt dat.etah}      &&    distribution in pseudo-rapidity of Higgs.
}
In addition, for ${\tt shtr>0}$, \HAWK\ also provides
\cpctable{{\tt dat.mhwide}}{
{\tt dat.mh}        &&    distribution in invariant mass of off-shell Higgs boson, centered around the Higgs-boson mass,\\
{\tt dat.mhwide}    &&    distribution in invariant mass of off-shell Higgs boson in a wide range.
}
For VBFH production, i.e.\ {\tt selprocess=0}, the following jet distributions are 
available,
\cpctable{{\tt dat.dphijjmaxpt}}{
{\tt dat.ptjmax1pt}  &&   distribution in transverse momentum of
leading jet $j_1$,\\
{\tt dat.ptjmax2pt}  &&   distribution in transverse momentum of subleading jet $j_2$,\\
{\tt dat.yjmax1pt}   &&   distribution in rapidity of leading jet $j_1$,\\
{\tt dat.yjmax2pt}   &&   distribution in rapidity of subleading jet $j_2$,\\
{\tt dat.dyjjmaxpt}  &&   distribution in rapidity difference of leading jets,\\
{\tt dat.dphijjmaxpt}&&   distribution in azimuthal angle difference of leading jets,\\
{\tt dat.mjjmaxpt}   &&   distribution in invariant mass of leading jets,
}
where the leading and subleading jets are ordered in $\pt$. 

For the Higgs-strahlung
processes, ${\tt selprocess>0}$, \HAWK\ provides
\cpctable{{\tt dat.ptv}}{
{\tt dat.ptv}       &&    distribution in transverse momentum of the vector boson ($\PW^{\pm}$ or $\PZ$)
}
and the following distributions concerning the final-state leptons
\cpctable{{\tt dat.dphilpH}}{
{\tt dat.ptlp}      &&    distribution in transverse momentum of $\Pl^+$,\\
{\tt dat.ylp}       &&    distribution in rapidity of $\Pl^+$,\\
{\tt dat.etalp}     &&    distribution in pseudo-rapidity of $\Pl^+$,\\
{\tt dat.ptmiss}    &&    distribution in transverse momentum of $\nu$ (= $p_{\mathrm{T,miss}}$),\\
{\tt dat.ymiss}     &&    distribution in rapidity of $\nu$,\\
{\tt dat.etamiss}   &&    distribution in pseudo-rapidity of $\nu$,\\
{\tt dat.dphilpH}   &&    distribution in azimuthal angle difference of $\Pl^+$ and Higgs
}
for $\PW^+\PH$ production. The analogous distributions are available for 
$\PW^-\PH$ production ({\tt lp} is replaced by {\tt lm} for the $\Pl^-$ 
distributions). For $\PZ\PH$ production with $\PZ$-boson decay into charged leptons, 
the distributions are of course available for the negatively and the positively charged
lepton. For $\PZ\PH$ production with $\PZ$-boson decay into neutrinos, the 
distributions for missing transverse momentum refer to the sum of the neutrino momenta.

{The relative EW corrections $\delta_{\mathrm{EW}}(\mathcal{O})$ for a bin-by-bin 
differential reweighting, as suggested in \refse{se:combination}, are obtained by 
the ratio of entries in the columns {\tt Monte Carlo average for virt+real ew cross-section} and \linebreak
{\tt Monte Carlo average for the Born cross-section} in each bin as provided in the text file for a 
given distribution. The contributions due to photons in the initial state are given in the column 
{\tt Monte Carlo average for incoming photon cross-section}.}

{The} histograms can be adapted, and other histograms can be included
by modifying the respective subroutine {\tt settings\_Hxx} in {\tt vbfh\_public.F} or
{\tt whzh\_public.F}, respectively.

For the histograms in \citere{Ciccolini:2007ec}
$10^9$ events were used, and the histograms were rebinned to 20 bins.

\subsection{Sample runs}

We provide several sample input files along with the corresponding results.
They can be found in the corresponding subdirectories of\linebreak {\tt HAWK-2.0/sampleruns}.
All sample runs (apart from {\tt input\_default}) use $10^7$ events and default
input parameters and cuts for the corresponding processes. They typically
take between 0.5 hours ({\tt testrun.Hnn}) to 2.5 hours ({\tt testrun.Hjj\_ofs})
(with the ifort compiler, with gfortran run times are typically a factor
2$-$3 larger). The following sample runs are available:\\[2ex]
{\tt input\_default}:\\
This file shows all input parameters with their default values and
provides a run with $10^6$ events for VBFH, $\Pp\Pp \to \PH+2\,\mathrm{jets}$, 
for on-shell Higgs boson, taking about 15 minutes (with the ifort compiler).
The output can be found in {\tt output\_default}.
The file {\tt input\_sample} is equivalent to {\tt input\_default}, making use
of the default settings in HAWK.\\[2ex]
{\tt input\_samplerun.Hjj}:\\*
sample run for VBFH, $\Pp\Pp \to \PH+2\,\mathrm{jets}$, for on-shell Higgs boson.\\[2ex]
{\tt input\_samplerun.Hjj.ofs}:\\*
sample run for VBFH, $\Pp\Pp \to \PH+2\,\mathrm{jets}$, for off-shell Higgs boson using an off-shell
propagator.\\[2ex]
{\tt input\_samplerun.Hmm}:\\*
sample run for $\Pp\Pp \to \PH\PZ \to \PH \Pmup \Pmum$, for bare muons and on-shell Higgs boson.\\[2ex]
{\tt input\_samplerun.Hnn}:\\*
sample run for $\Pp\Pp \to \PH\PZ \to \PH \Pane \Pne$, for on-shell Higgs boson.\\[2ex]
{\tt input\_samplerun.Hmn}:\\*
sample run for $\Pp\Pp \to \PH\PW^+ \to \PH \Pmup \Pne$, for bare
anti-muon and on-shell Higgs boson.\\[2ex]
{\tt input\_samplerun.Hnm}:\\*
sample run for $\Pp\Pp \to \PH\PW^- \to \PH \Pane \Pmum$, for bare muon and on-shell Higgs boson.\\[2ex]
{\tt input\_samplerun.Hmn.undet\_lepton}:\\*
sample run for $\Pp\Pp \to \PH\PW^+ \to \PH \Pmup \Pne$, for on-shell Higgs boson and
undetected bare anti-muon.\\[2ex]
{\tt input\_samplerun.Hjj.ac}:\\*
sample run for VBFH, $\Pp\Pp \to \PH+2\,\mathrm{jets}$, for on-shell Higgs boson with anomalous HVV couplings.\\[2ex]
{\tt input\_samplerun.Hmn.ac}:\\*
sample run for $\Pp\Pp \to \PH\PW^+ \to \PH \Pmup \Pne$, for bare anti-muon and on-shell
Higgs boson with anomalous HVV couplings.

\section{Conclusions}
\label{se:conclusion}

The Monte Carlo program \HAWK\ provides predictions for Higgs-boson production
via weak vector-boson fusion or Higgs strahlung ($\PV\PH$ production) at hadron colliders,
including the full set of next-to-leading order corrections of the strong and electroweak
interactions. 
In $\PV\PH$ production, leptonic decays of the weak gauge boson
$\PV=\PW,\PZ$ and the corresponding off-shell effects are included at this level of accuracy.
Off-shell effects of the Higgs boson can be simulated as well, but only
for vector-boson fusion a decay into a pair of massless singlets is
supported in the current version.

Predictions are provided in the framework of the Standard Model, optionally extended
by anomalous Higgs--gauge-boson interactions. 
Further generalizations of the program to models with extended Higgs sectors
are planned for future updates of \HAWK.

\subsection*{Acknowledgements}

The work of A.D.  was partially supported by the Bundesministerium f\"ur
Bildung und Forschung (BMBF) under under contract no. 05H12WWE.





\bibliographystyle{elsarticle-num}
\bibliography{<your-bib-database>}







\end{document}